\documentclass[twocolumn,showpacs, prl]{revtex4}

\usepackage{graphicx}
\usepackage{dcolumn}
\usepackage{bm}

\begin{document}
\title{Valley susceptibility of an interacting two-dimensional electron system}
\date{\today}
\author{O. Gunawan}
\author{Y. P. Shkolnikov}
\author{K. Vakili}
\author{T. Gokmen}
\author{E. P. De Poortere}
\author{M. Shayegan}

\affiliation{Department of Electrical Engineering, Princeton University, Princeton, NJ 08544}

\begin{abstract}
We report direct measurements of the valley susceptibility, the change of valley population in
response to applied symmetry-breaking strain, in an AlAs two-dimensional electron system. As the
two-dimensional density is reduced, the valley susceptibility dramatically increases relative to
its band value, reflecting the system's strong electron-electron interaction. The increase has a
remarkable resemblance to the enhancement of the spin susceptibility and establishes the analogy
between the spin and valley degrees of freedom.
\end{abstract}


\maketitle

\begin{figure*}
\includegraphics[width=150mm]{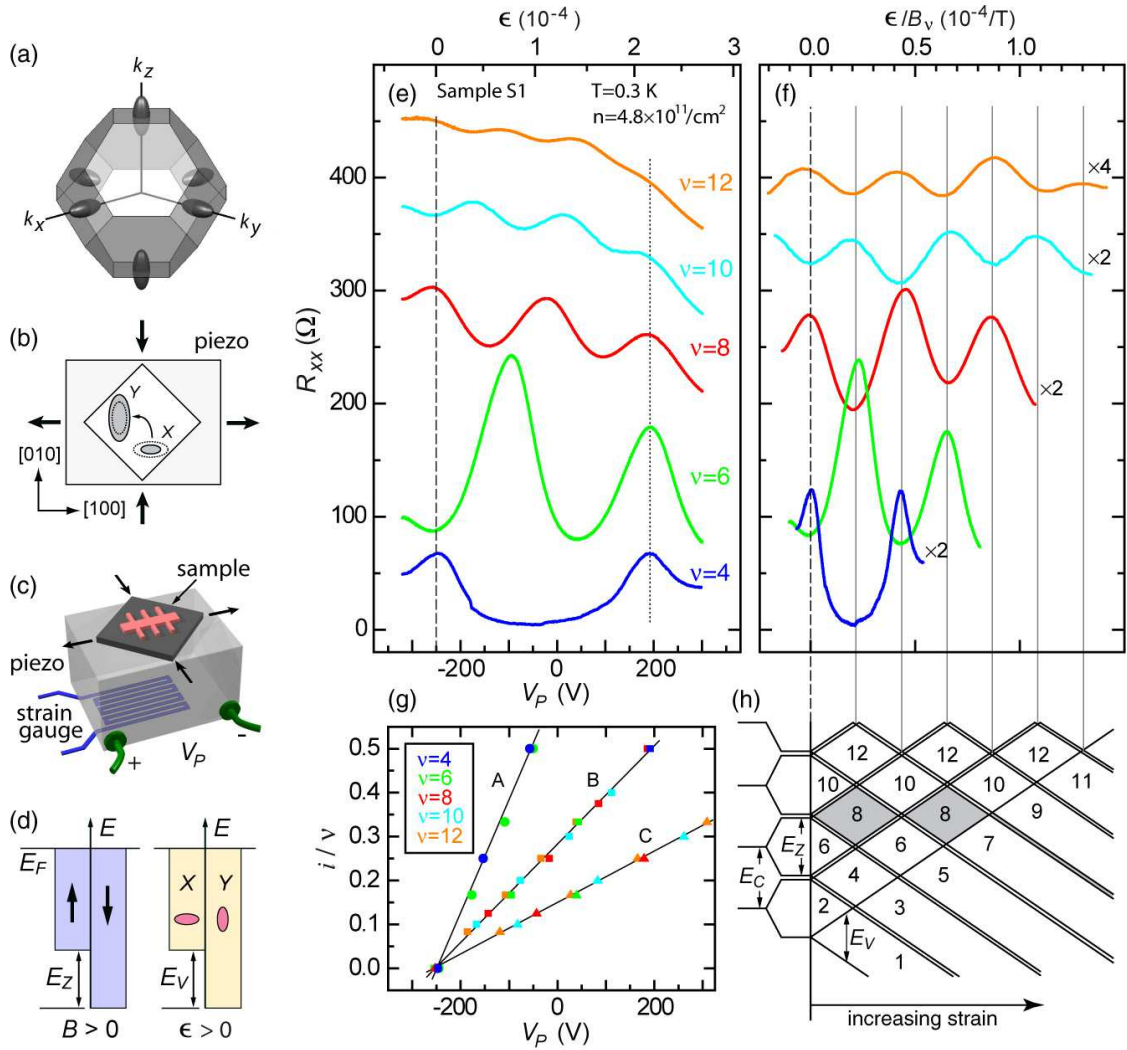}
\caption{ (a) The first Brillouin zone and conduction band constant energy surfaces for bulk AlAs.
(b) Schematic diagram showing that electrons are transferred from the $X$ to the $Y$ valley upon
the application of strain while the total 2DES density is independent of strain. (c) Experimental
setup for the valley susceptibility measurements. (d) Energy diagram showing the spin ($E_Z$) and
valley ($E_V$) subband splittings with applied magnetic field or strain, respectively. (e) Sample
resistance, $R_{xx}$, vs the piezoelectric actuator bias, $V_P$ at various filling factors $\nu$.
The traces are offset for clarity. The dashed and dotted lines respectively mark $V_P$ at which
$X$ and $Y$ are equally populated, and $V_P$ beyond which all the electrons are transferred to the
$Y$ valley. The top axis shows the strain as measured by the strain gauge. (f) Same data as in (e)
but now plotted vs a normalized horizontal (top) axis $\epsilon/B_{\nu}$, showing that the
oscillations have the same period. The traces are offset for clarity and, for $\nu$=8, 10 and 12,
a smooth, parabolic background is subtracted from the traces to highlight the oscillations. (g)
Plot of the coincidence index, $i$ normalized by $\nu$, showing linearity with $V_P$. From the
slopes of the lines fitted through the data points we obtain $\chi_v$, and the $i=0$ intercepts
give the value of $V_P$ at which the $X$ and $Y$ valleys are equally populated ($\epsilon=0$).
Three data sets A, B, and C correspond to densities of: 2.8, 4.8 and 7.6$\times$10$^{11}$
cm$^{-2}$, respectively. (h) Schematic energy fan diagram showing the relevant energies: the
cyclotron ($E_C$), Zeeman ($E_Z$) and valley splitting ($E_V$).}
 \label{FigExp}
\end{figure*}

Currently, there is considerable interest in controlled manipulation of electron spin in
semiconductors. This interest partly stems from the technological potential of spintronics,
namely, the use of carrier spin to realize novel electronic devices. More important, successful
manipulation of spins could also impact the more exotic field of quantum computing since many of
the current proposals envision spin as the quantum bit (qubit) of information
\cite{LossPRA98,KaneNat98,PettaSci05}. Here we describe measurements of another property of
electrons, namely their valley degree of freedom, in a semiconductor where they occupy multiple
conduction band minima (valleys) [Fig.~\ref{FigExp}(a)]. Specifically, for a two-valley,
two-dimensional electron system (2DES) in an AlAs quantum well, we have determined the "valley
susceptibility", $\chi_{v}$, {\it i.e.}, how the valley populations respond to the application of
symmetry-breaking strain. This is directly analogous to the spin susceptibility, $\chi_{s}$, which
specifies how the spin populations respond to an applied magnetic field [Figs.\ 1(d)]. Our data
show that $\chi_{v}$ and $\chi_{s}$ have strikingly similar behaviors, including an
interaction-induced enhancement at low electron densities. The results establish the general
analogy between the spin and valley degrees of freedom, implying the potential use of valleys in
applications such as quantum computing. We also discuss the implications of our results for the
controversial metal-insulator transition problem in 2D carrier systems.

It is instructive to describe at the outset the expressions for the {\it band} values of spin and
valley susceptibilities $\chi_{s,b}$ and $\chi_{v,b}$ \cite{BandParameter}. The spin
susceptibility is defined as $\chi_{s,b} = d{\Delta n}/{dB} = g_b \mu_{B}\rho/2$, where $\Delta n$
is the net spin imbalance, {\it B} is the applied magnetic field, ${\it g_b}$ is the band
Land$\acute{e}$ $g$-factor, and $\rho$ is the density of states at the Fermi level. Inserting the
expression $\rho = m_b/\pi\hbar^2$ for 2D electrons, we have $\chi_{s,b}$ =
($\mu_B/2\pi\hbar^2)g_b m_b$, where $m_b$ is the band effective mass. In analogy to spin, we can
define valley susceptibility as $\chi_{v,b} = d{\Delta n}/d\epsilon= \rho E_{2,b}$
$=(1/\pi\hbar^2)m_b E_{2,b}$, where $\Delta n$ is the difference between the populations of the
majority and minority valleys, $\epsilon$ is strain, and $E_{2,b}$ is the conduction band
deformation potential \cite{strain}. In a Fermi liquid picture, the interparticle interaction
results in replacement of the parameters $m_b$, $g_b$, and $E_{2,b}$ \cite{BandValues} by their
normalized values $m^*$, $g^*$, and $E_2^*$. Note that $\chi_s\propto m^*g^*$ and $\chi_v\propto
m^*E_2^*$.

Our experiments were performed on a high-mobility 2DES confined to an 11 nm-thick,
modulation-doped layer of AlAs grown by molecular beam epitaxy on a (001) GaAs substrate
\cite{PoortereAPL02}. We studied two samples, each patterned in a standard Hall bar mesa aligned
with the [100] crystal direction [Figs.\ 1(b) and (c)]. Using a metal gate deposited on the
sample's surface we varied the 2DES density $n$, between $2.5$ and 9.5$\times$10$^{11}$ cm$^{-2}$.
The magneto-resistance measurements were performed in a liquid $^3$He system with a base
temperature of 0.3 K.

In bulk AlAs electrons occupy three (six half) ellipsoidal conduction valleys at the six
equivalent X-points of the Brillouin zone [Fig.~\ref{FigExp}(a)]. For the 11 nm-wide quantum well
used in our experiments, only the valleys with their major axes along [100] and [010] are occupied
\cite{PoortereAPL02}; we refer to these as the {\it X} and {\it Y} valleys, respectively. The
application of symmetry-breaking strain along [100] and [010] splits the energies of the {\it X}
and {\it Y} valleys, transferring charge from one valley to the other [Fig.~\ref{FigExp}(b)]. To
apply tunable strain we glued the sample to one side of a piezoelectric (piezo) stack actuator
\cite{ShayeganAPL03}. The piezo polling direction is aligned along [100] as shown in
Fig.~\ref{FigExp}(c). When bias $V_P$ is applied to the piezo stack, it expands (shrinks) along
[100] for $V_P> 0$ ($V_P < 0$) and shrinks (expands) in the [010] direction. We have confirmed
that this deformation is fully transmitted to the sample and, using metal strain gauges glued on
the opposite side of piezo [Fig.~\ref{FigExp}(c)], have measured its magnitude in both [100] and
[010] directions \cite{ShayeganAPL03, PiezoCalib}.

We determined $\chi_v$ via different techniques. The first involves measuring, at $B = 0$,
sample's resistance ($R_{xx}$), along [100] as we induce tensile (compressive) strain and the
electrons are transferred to the $Y$ ($X$) valley [Fig.~\ref{FigExp}(b)]. Because of the smaller
(larger) effective mass of electrons in this valley along [100], $R_{xx}$ decreases (increases) as
a function of $V_P$ and saturates once all the electrons are transferred to the $Y$ ($X$) valley
\cite{ShkolnikovAPL04}. We can use the onset of the saturation as a signature of the full electron
transfer and, from the corresponding $\epsilon$, determine $\chi_v$ \cite{ChiValleyDepop}. Our
second technique is based on measuring the valley densities at a fixed $V_P$ from the Shubnikov-de
Haas oscillations of $R_{xx}$ as a function of perpendicular magnetic field, $B_{\perp}$. The
frequencies of these oscillations are proportional to the densities of the two valleys and provide
a direct measure of $\chi_v$ \cite{ChiSdH}. We quote data obtained from these two techniques later
in the paper \cite{DepopSdH}; here we would like to focus on a powerful variation of the second
technique which also directly measures $\chi_v$.

The technique is based on the "coincidence" of the 2DES's energy levels at the Fermi energy
($E_F$) and entails monitoring oscillations of $R_{xx}$ as a function of  $\epsilon$ at a {\it
fixed} $B_{\perp}$ \cite{ChiCoincidence}. Examples of such data are shown in Fig.~\ref{FigExp}(e)
for $n=$4.8$\times$10$^{11}$ cm$^{-2}$ and various values of $B_{\perp}$ corresponding to even
Landau level (LL) filling factors $\nu$ = 4, 6,..., 12. Clear oscillations of $R_{xx}$ as a
function of $V_P$ are seen. These oscillations come about because the strain-induced
valley-splitting causes pairs of quantized energy levels of the 2DES in $B_{\perp}$ to cross at
$E_F$.  As schematically shown in the fan diagram of Fig.~\ref{FigExp}(h), there are three main
energies in our system. The field $B_{\perp}$ quantizes the allowed energies into a set of LLs,
separated by the cyclotron energy, $E_C = \hbar eB_{\perp}/m^*$. Because of the electron spin,
each LL is further split into two levels, separated by the Zeeman energy, $E_Z = \mu_B g^*B$. In
the absence of in-plane strain, each of these energy levels in our 2DES should be two-fold
degenerate. By straining the sample, we remove this degeneracy and introduce a third energy, the
valley-splitting, $E_V = \epsilon E_2^*$, which increases linearly with strain, as illustrated in
Fig.~\ref{FigExp}(h). When an integer number of the quantized energy levels of a 2DES are exactly
filled, $E_F$ falls in an energy gap separating adjacent levels (see, e.g. the shaded area of
Fig.~\ref{FigExp}(h) for the case of $\nu = 8$) and, at low temperatures, $R_{xx}$ exhibits a
local minimum. As is evident in Fig.~\ref{FigExp}(h), however, at certain values of $\epsilon$,
the energy levels corresponding to different valley- and spin-split LLs coincide at $E_F$. At such
"coincidences", the $R_{xx}$ minimum becomes weaker or disappears altogether
\cite{ChiCoincidence}. From the fan diagram of Fig.~\ref{FigExp}(h) we expect the
weakening/strengthening of the $R_{xx}$ minimum to be a periodic function of $\epsilon$, or $V_P$,
since it happens whenever $E_V$ is an even multiple of $E_C$. Therefore we can directly measure
$E_V$ (in units of $E_C$) and determine the valley susceptibility: $\chi_v= 4eB_{\nu}/h\Delta
\epsilon$, where $\Delta \epsilon$ is the period of the $R_{xx}$ oscillations and $B_{\nu}$ is
$B_{\perp}$ of the filling factor $\nu$ at which the oscillations are measured.

Before we discuss the measured values of $\chi_v$ as a function of density, we point out several
noteworthy features of Figs.\ 1(e)-(g) data. First, because of finite residual stress during the
cooling of the sample and the piezo, we need a finite, cooldown-dependent $V_P$ to attain the
zero-strain condition in our experiments; this is about -250 V for the experiments of Figs.\
1(e)-(g) and is marked by a dashed vertical line in Fig.~\ref{FigExp}(e). The data of
Fig.~\ref{FigExp}(e) themselves allow us to determine the zero-strain condition: at $V_P$ = -250
V, $R_{xx}$ at $\nu$ = 4, 8, and 12 is at a (local) maximum while at $\nu$ = 6 and 10 it is at a
(local) minimum. This behavior is consistent with the energy diagram of Fig.~\ref{FigExp}(h) when
no strain is present. Second, the period of the $R_{xx}$ oscillations as a function of $\epsilon$
is larger at smaller $\nu$ (larger $B_{\perp}$). In Fig.~\ref{FigExp}(f) we plot $R_{xx}$ as a
function of $\epsilon$ normalized by $B_{\nu}$. It is clear in this plot that the oscillations are
in-phase at $\nu$ = 4, 8, and 12, and that these are $180^{\circ}$ out-of-phase with respect to
the oscillations at $\nu$ = 6 and 10. This is consistent with the simple fan diagram of
Fig.~\ref{FigExp}(h) \cite{FanDiagram}. To quantify this periodicity and also compare the data for
different densities, in Fig.~\ref{FigExp}(g) we show plots of the coincidence "index",
$i=E_V/E_C$, divided by $\nu$, vs $V_P$; the indices are obtained from the positions of maxima and
minima of the $R_{xx}$ oscillations and the index at the zero-strain condition is assigned to be
zero. The plots in Fig.~\ref{FigExp}(g) show that the valley polarization is linear in $\epsilon$.
This linearity means that $\chi_v$ is independent of valley polarization or $\nu$ (for $\nu \geq
4$) within our experimental uncertainty \cite{ChiSNote}. Third, in Fig.~\ref{FigExp}(e) $R_{xx}$
at all $\nu$ shows a maximum at a particular value of $\epsilon$ (dotted vertical line at $V_P
\sim200$ V). This maximum marks the last "coincidence" of the energy levels. Beyond this value of
$\epsilon$, the minority valley is completely depopulated and there are no further oscillations.
We have confirmed this statement at lower sample densities where our available range of $V_P$
allows us to go well beyond the valley depopulation $\epsilon$.

\begin{figure}
\includegraphics[scale=1.05]{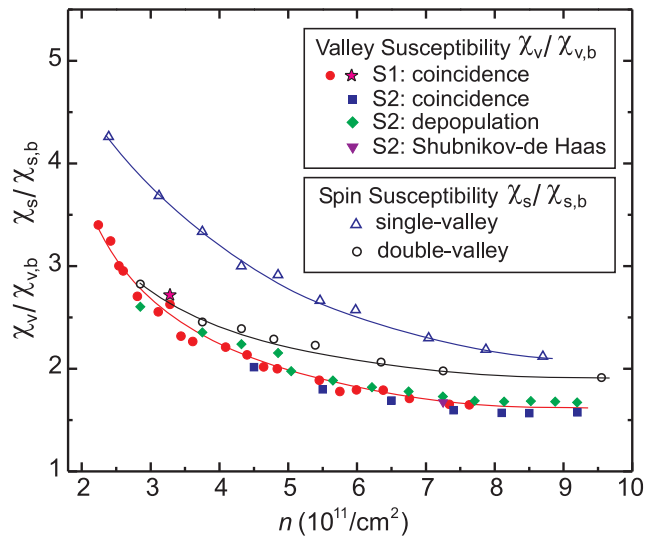} \caption{
Valley and spin susceptibilities, normalized to their band values, vs density. $\chi_v$ was
measured for two samples, S1 and S2, and using various techniques, as indicated. $\chi_s$ data
were measured in several AlAs 2DES samples, including S2, and are shown for the case when the $X$
and $Y$ valleys are equally occupied (open circles), and for when only one valley is occupied
(open triangles). Lines are guides to the eye.}
 \label{FigChi}
\end{figure}

Figure~\ref{FigChi} summarizes our main result, namely, the measured $\chi_v$ as a function of
density. $\chi_v$ is enhanced at all accessible densities and particularly at lower $n$. This
enhancement is in sharp contrast to the non-interacting picture where $\chi_v$ should maintain its
band value at all electron densities, and betrays the strong influence of electron-electron
interaction in this system.

A comparison of the measured $\chi_v$ with the $spin$ susceptibility for similar samples
\cite{ShkolnikovPRL04} supports this conjecture. As seen in Fig.~\ref{FigChi}, $\chi_s$ and
$\chi_v$ show a remarkably similar enhancement. The increase of $\chi_s$ as $n$ is lowered is well
established theoretically \cite{AttaccalitePRL02} and experimentally for various 2DESs
\cite{OkamotoPRL99,ShashkinPRL01,ZhuPRL90,PudalovPRL88, VakiliPRL04, ShkolnikovPRL04}. It occurs
because of the increasing dominance of the Coulomb energy over the kinetic (Fermi) energy as $n$
is decreased. To minimize its Coulomb energy, the 2DES tends to more easily align the electrons'
spins so that they would spatially stay farther apart from each other. A similar mechanism is
likely leading to the $\chi_v$ enhancement. This is not unexpected, if we regard valley and spin
as similar degrees of freedom. Indeed, in the standard effective mass approximation, where
electrons near a semiconductor's conduction band minima are treated as quasi-particles with
re-normalized ({\it i.e.}, $band$ values) of parameters such as (the effective) mass and
$g$-factor but are otherwise free, spin and valley are equivalent, SU(2) symmetric degrees of
freedom. Interaction would then be expected to affect the spin and valley susceptibilities in a
similar fashion. Interaction-induced valley-splitting, akin to exchange-induced spin-splitting, is
in fact, known to occur in multi-valley 2DESs in the presence of $B_{\perp}$
\cite{ShkolnikovPRL02}. Evidence was also recently reported for interaction-induced quantum Hall
valley-Skyrmions, in direct analogy with spin-Skyrmions \cite{ShkolnikovPRL05}.

Despite the similarities between the valley and spin degrees of freedom, there remains a puzzle.
As shown in Fig.~\ref{FigChi}, within our experimental uncertainty, there is no spin polarization
dependence of $\chi_v$, although there is a valley polarization dependence of $\chi_s$
\cite{ShkolnikovPRL04}. Our $\chi_v$ measurements are performed primarily at fixed $\nu\geq 4$ in
a regime where the electrons are not fully spin polarized. But we have also measured $\chi_v$ at
$\nu=4$ in tilted magnetic fields, where the 2DES is fully spin polarized, and it agrees closely
with $\chi_v$ in the un-polarized regime (see "star" data point in Fig.~\ref{FigChi}). However, as
seen in Fig.~\ref{FigChi}, $\chi_s$ in a single valley system is larger than in the two valley
case \cite{ShkolnikovPRL04}. This disparity may point to a fundamental difference between the
valley and spin degrees of freedom, possibly reflecting the inadequacy of the effective mass
approximation when one is dealing with interaction between electrons in different, {\it
anisotropic} valleys located at different points of the Brillouin zone.

We close by making some general remarks. First, utilizing the spin degree of freedom to make
functional devices (spintronics), or as a qubit for quantum computation, has been of much interest
lately \cite{LossPRA98,KaneNat98,PettaSci05}. The valley degree of freedom may provide an
alternative for such applications. For example, it may be possible to encode quantum information
into the "valley state" of a two-level quantum dot. This type of qubit could potentially have much
longer coherence times than those of the spin-based GaAs qubits where the electron spin coherence
times suffer from a coupling of the electron and nuclear spins (see, e.g., Ref.
\cite{PettaSci05}). The results reported here demonstrate that the valley populations can be
modified $and$ monitored, and provide values for one of the most basic parameters of the system,
namely its (valley) susceptibility. Second, the spin susceptibility of an interacting 2DES is
expected to eventually diverge when $n$ is lowered below a critical value at which the system
attains a ferromagnetic ground state. There are also transitions to ferromagnetic quantum Hall
states in samples with appropriate parameters. We believe that, under favorable conditions, e.g.,
at very small $n$ or in the quantum Hall regime, "valley ferromagnetism" phenomena should also be
prevalent. Finally, it is believed that there exists an intimate relationship between spin
polarization and the 2D metal-insulator transition (MIT), and some have even suggested that the
spin susceptibility diverges at the zero-field MIT  critical density \cite{ShashkinPRL01}. The
similarity of spin and valley degrees of freedom suggests that the latter should play a role in
the MIT problem as well. We have studied the MIT in the present system and our preliminary results
indeed indicate that the 2DES can be driven to an insulating phase by increasing its valley
polarization. A full account of our findings will be reported elsewhere.

We thank the NSF and ARO for support, and R. Bhatt, S.A. Lyon, E. Tutuc, and R. Winkler for
illuminating discussions.


\begin{references}
\bibitem{BandParameter}
These "band" values differ in general from the values for electrons in vacuum because of the
electron's interaction with the host material's atoms.
\bibitem{strain}
We define strain as $\epsilon=\epsilon_{[100]}-\epsilon_{[010]}$. With this definition, the
splitting in $X$ and $Y$ valley energies is equal to $\epsilon E_{2,b}$.
\bibitem{ShkolnikovAPL04}
Y.\ P.\ Shkolnikov {\it et al.}, Appl.\ Phys.\ Lett.\ {\bf 85}, 3766 (2004); note, however that
the applied strain was overestimated in that work because of a miscalibration.
\bibitem{ChiValleyDepop}
Determining $\chi_v$ from the onset of full valley polarization is analogous to measuring $\chi_s$
for 2DESs from the onset of full spin polarization using parallel magnetic field.
\bibitem{DepopSdH}
In Fig.~\ref{FigChi} we refer to these as "depopulation" and "Shubnikov-de Haas" data.
\bibitem{FanDiagram}
This fan diagram is qualitatively correct in the density range of interest where $0.9 \leq E_Z/E_C
\leq 1.4 $.
\end{references}
\end{document}